# Formal Verification of Full-Wave Rectifier: A Case Study

Kusum Lata , H S Jamadagni

**Abstract** — *We present a case study of formal verification of full-wave rectifier for analog and mixed signal designs. We have used the Checkmate tool from CMU [1], which is a public domain formal verification tool for hybrid systems. Due to the restriction imposed by Checkmate it necessitates to make the changes in the Checkmate implementation to implement the complex and non-linear system. Full-wave rectifier has been implemented by using the Checkmate custom blocks and the Simulink blocks from MATLAB from Math works. After establishing the required changes in the Checkmate implementation we are able to efficiently verify the safety properties of the full-wave rectifier.* [1]

## I. INTRODUCTION

Formal verification has become part of the design process of digital circuits; its application to analog and mixed-signal (AMS) design is only in its infancy. This is mainly due to the fact that the mathematical models for analog and mixed-signal circuits are very different from the discrete, finite-state transition systems which are used for the verification of digital systems. These models for AMS designs are based on continuous dynamical systems governed by differential equations and their verification calls for different techniques, like those developed in the analysis of hybrid systems.

Mathematically, the behavior of analog circuits can be described by continuous variables and a set of differential equations, whereas discrete variables and switching-modes are also used for modeling the mixed-signal circuits. Thereby, analog and mixed-signal circuits are hybrid system in nature. The analysis of continuous and mixed discrete-continuous systems is inherently difficult and many different abstractions in combination with dedicated verification techniques are currently being investigated by researchers. Several attempts [2, 3, 4, 5] have been made to apply the formal verification techniques of hybrid systems in the context of the formal verification of analog and mixed-signal circuits.

In this paper we present a case study of formal verification of full wave rectifier (FWR) for the analog and mixed signal designs. The FWR is implemented in Checkmate along with Simulink blocks for simulation as well as for the formal analysis. Checkmate is a public domain tool from CMU [1], for formal verification of hybrid systems. We have formulated the safety properties for the full wave rectifier. Authors have reported number of tools in [6], for the simulation as well as formal verification of FWR. But in this author has used the differential equations for defining the dynamics of the circuit or the state-space approach. But in several cases this route becomes very restrictive so we need to have some mechanism for defining the other complex systems too so that these formal verification techniques become feasible to the complex systems having non-linear components in it, Where we do not have the exact differential equations for defining the behavior of the system or state space equations. But they are easily simulated in several simulation frameworks.

This paper is organized as follows: In Section II we give a brief overview of Checkmate. The proposed approach to formal verification in Checkmate using simulation traces is discussed in Section III. In section IV, we illustrate our approach on the case study (i.e. FWR) along with results. We conclude with some pointers to future work in Section V.

## II. FORMAL VERIFICATION IN CHECKMATE

Checkmate is a public domain formal verification tool for hybrid systems from CMU [1]. Checkmate is built on top of the SimuLink/StateFlow framework (SSF) from Math Works based on numerical simulation. Our choice of Checkmate for formal analysis is driven by the fact that SSF is a widely used design and validation framework for hybrid systems in the industrial context. System level models of hybrid systems are constructed using a large class of generic blocks in SSF and then validated using a discrete time domain dynamic simulation approach. For formal analysis, Checkmate accepts a hybrid system modeled as a restrictive hybrid automaton known as polyhedral invariant hybrid automata (PIHA) [1]. This necessitates transformation of the general SSF model into the restrictive PIHA model using a subset of SSF blocks accepted by Checkmate. Simulation traces generated from this model are used by the formal engine in Checkmate to carry out formal analysis.

In several situations this route to formal analysis is highly limiting. For example, implementation models of the hybrid systems such as mixed signal design are not amenable to formal analysis in Checkmate, in case they have dynamic components described with a system of strongly non-linear differential-algebraic equations. However, these are easily simulated in frameworks such as, Simulink/State Flow, SABER or H-SPICE, and hence, should be amenable to formal analysis. Our approach attempts to bridge this gap in applying formal analysis in Checkmate to the simulation traces generated from a real implementation model. Our approach utilizes the simulation traces generated from the implementation model in SSF, and the thresholds on the design variables in the formal verification environment of Checkmate.

For formal analysis, Checkmate accepts a hybrid system modeled as a restrictive hybrid automaton known as

Kusum Lata[1] and H S Jamadagni are with the Center for Electronics Design and Technology, Indian Institute of Science, Bangalore, India (e-mail: {lkusum,hsjam}@cedt.iisc.ernet.in).

polyhedral invariant hybrid automata (PIHA) [1]. Checkmate constructs a flow-pipe of the trajectories over time originating from the given set of initial states. This flow pipe represents the set of reachable points in the vector space of state variables of the hybrid system. It then approximates this flow-pipe with overlapping linear polyhedrons [1, 7].

In Checkmate, three important custom SSF blocks (i.e. Switched Continuous System Block (SCSB), Polyhedral Threshold Block (PTHB) and Finite State Machine Block (FSMB)) are available. A hybrid system is implemented using these blocks from Checkmate along with few other blocks in SSF. A SCSB block is used to define the system continuous dynamics in terms of first order differential equations. A PTHB block generates events whenever the system crosses a specified threshold described in terms of a linear constraint. This generated event is used as an input to the FSMB block to trigger transitions from one state to another. Based on the sink state of a transition edge that is reached, the SCSB block generates the continuous state trajectory corresponding to that state. Checkmate follows three steps to verify hybrid systems. In first step it allows one to *simulate* the hybrid system. Checkmate models can be simulated like any other general SSF model. In the next step, *Explore* phase in Checkmate checks whether each simulated trajectory starting with different initial states (chosen from a minimal set of points in the initial continuous set) satisfies a given formal property specified as an ACTL formula. It informs the user in case of a violation. The *Verify* phase uses these trajectories to construct the flow pipe and its approximation (using polyhedrons). Prior to initiating construction of the flow-pipe, it checks each block in the model for compliance with Checkmate requirements. Formal analysis for fail safe behavior based on computational geometry algorithms is carried out after this for each property. *Verify* completes one iteration of the verification process to determine whether or not the system satisfies the specification. The user is informed of the outcome of this verification effort. If the verification concludes, the program terminates; else, Checkmate attempts to refine the set of initial continuous set by partitioning it, and then iterates with respect to a subset from this partition.

## III. FORMAL VERIFICATION USING SIMULATION TRACES IN CHECKMATE

In general, analog and mixed signal designs are carried out first by modeling it in SSF. The system level model is constructed by using a large class of generic blocks and then validating it through built in numerical simulation algorithms available in SSF. Due to the restriction imposed by Checkmate it becomes very difficult to implement the circuits and systems having non-linear dynamics in the Checkmate since it does not allow any input from out side as well as any output from the other blocks. We need to have the mechanism in the tool so that it accepts the output from another block other than Checkmate custom block and it should be able to accept the external input. Since Checkmate imposes so many restrictions therefore it necessitates for making the changes in the implementation itself. To accommodate the external voltage input source and the input of the PTHB blocks from the Simulink subsystems we have made the following changes: 1) PTHB blocks accepts only the output from SCSB blocks but in case of nonlinear behavior or the external input from other blocks PTHB blocks should be able to accept these inputs. Therefore it becomes necessary to make the changes in the in-built Checkmate function. This function traces back for the input to the PTHB blocks and allows only to the SCSB blocks output. After making the appropriate changes in this function it allows us to take the required inputs as discussed above. 2) Checkmate uses the user defined function from the SCSB block to do the formal analysis. It solves these ordinary differential equations from the user defined functions, with the help of ODE 45 solver. Since in several cases we do not have the ordinary differential equations, so we are not using this feature of SCSB block. Instead, we have implemented the continuous dynamics with the help of Simulink Blocks and using the output of these blocks as an input to the PTHB blocks. For this we have created a subsystem in Simulink which bypasses the SCSB output and takes the output of these blocks as an input to the PTHB Blocks. SCSB block is also needed to have the other important parameters which are needed for the formal analysis from this block. For example the values for initial continuous set (ICS), analysis region (AR) etc. are needed for the formal verification in Checkmate. Presence of the SCSB block enables Checkmate to precede with the desired formal analysis, as its compliance checking phase is done on this block. 3) In Checkmate, in-built function *simulate_points.m* uses the ODE45 solver to solve the user defined differential equations of the system. Since we are using the simulation traces from the subsystem output as an input to the PTHB block. So for that we are not using this ODE45 solver, instead we are using "sim" command for calling the simulation of the Checkmate model. Then it stores the output in the output variables and these variables values are used for the formal analysis in the Checkmate formal engine.

## IV. CASE STUDY: FULL-WAVE RECTIFIER

For the formal verification of analog and mixed signal designs the full wave rectifier has been simulated and formally verified as a case study. Full wave rectifier shown in figure (1), is used to obtain a constant voltage source starting from the sinusoidal one [2]. It consists of two diodes D1 and D2, a resistor R and a capacitor C and a voltage source with voltage $V_{in}$. In this circuit, when $V_{in} > 0$, diode D1 is in forward biased while D2 is reverse biased. When $V_{in} < 0$, diode D2 is forward biased and D1 is reverse biased. In both cases the current flows in the load in the same direction.

In [6], authors have implemented the full wave rectifier for the simulation purpose in Checkmate. But they have not reported any formal verification for that. Here we have reported the simulation as well as formal verification of the FWR with the modified Checkmate.

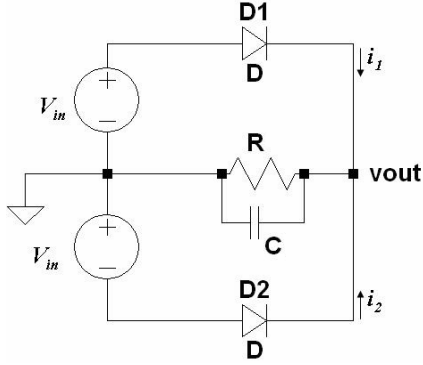

Figure 1: A Full Wave Rectifier Circuit

Checkmate model for the FWR is shown in figure 2. We are able to verify the safety properties for the FWR.

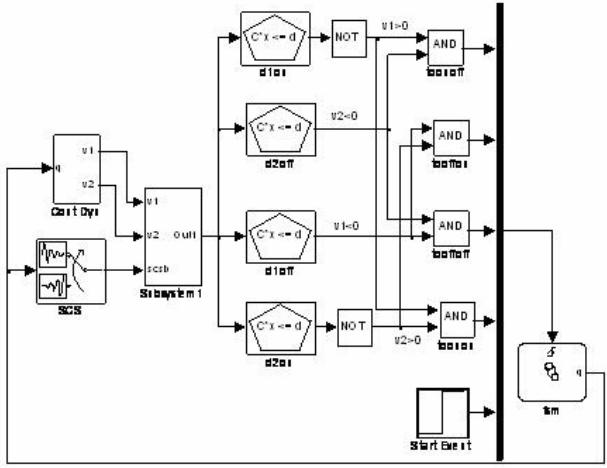

Figure 2: Checkmate model of the Full-Wave Rectifier

In figure 2, Cont Dyn and Subsystem1 are the two blocks which have been used to implement the dynamics of the FWR. Then the output of the Subsystems1 goes to the PTHB blocks as an input for carrying out the simulation as well as formal verification in Checkmate. Figure 3 shows the hybrid automata representing the FWR. In this automaton there are four states, which represent the different working condition combinations of the two diodes. In all the four cases, the continuous dynamics (has been taken from [6]) for the voltages is described by the following equations:

$$v_{in} = \sin(2\pi f t)$$
$$\dot{v}_{out} = -\frac{v_{out}}{RC} + \frac{i_1 + i_2}{C}$$
$$v_1 = v_{in} - v_{out}$$
$$v_2 = -v_{in} - v_{out}$$

The dynamics for the currents $i_1$ and $i_2$ and the invariant conditions for each state are as follows:

**OnOn:** When both the diodes are on, the continuous dynamics is described by the additional equations:
$$i_1 = v_1/R_f$$
$$i_2 = v_2/R_f$$
and the invariant is $v_1 \geq 0$ & $v_2 \geq 0$

**OnOff:** When D1 is on and D2 is off, the continuous dynamics is described by the additional equations:
$$i_1 = v_1/R_f$$
$$i_2 = -I_0$$
and the invariant is $v_1 \geq 0$ & $v_2 < 0$.

**OffOn:** When D2 is on and D1 is off, the continuous dynamics is described by the additional equations:
$$i_1 = -I_0$$
$$i_2 = v_2/R_f$$
and the invariant is $v_1 < 0$ & $v_2 \geq 0$.

**OffOff:** When both the diodes are off, the continuous dynamics is described by the additional equations:
$$i_1 = -I_0$$
$$i_2 = -I_0$$
and the invariant is $v_1 < 0$ & $v_2 < 0$.

The hybrid automaton for the FWR is shown in figure (3) with the above given four diode states.

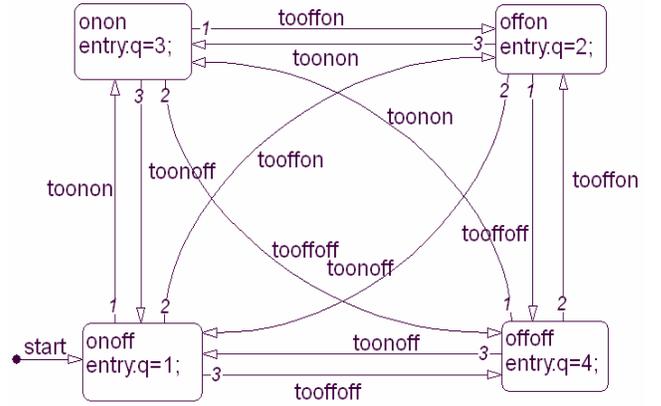

Figure 3: State Transition Graph for FWR Simulation in Checkmate

The simulation results are shown in figure (4). Figure 5 shows the State Transition graph for formal verification in Checkmate. In figure (5), "avoid" state is used for the verification purpose.

This state is used to generate the ACTL specification for the formal verification in Checkmate. It generates the following ACTL formula for defining the specification in Checkmate:

*"(AG ~ out_of_bound) & (AG ~ fsm == avoid)"*

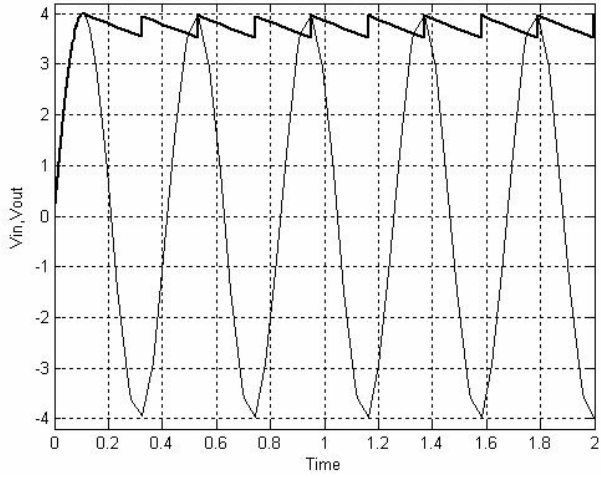

Figure 4: Checkmate simulation results of the FWR

Where out_of_bound defines the boundary of the location. We verify the following two basic properties (for safety) of the FWR:

**P1.** The voltage Vout is never negative.
**P2.** For a given $V_{in} = A\sin(2\pi ft)V$ and the initial condition Vout (0) = 4V, Vout (t) does not drop below a threshold.

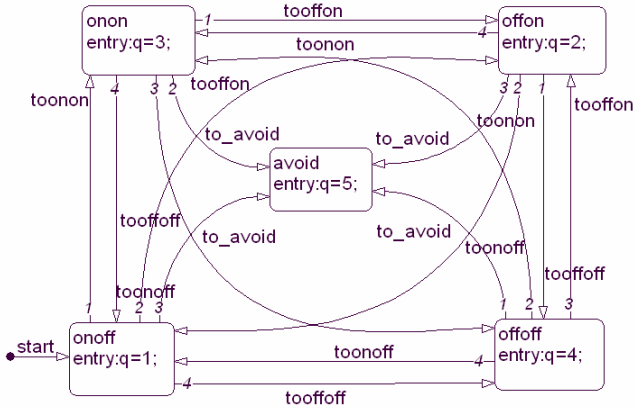

Figure 5: State Transition Graph for FWR Verification in Checkmate

After verifying each property, Checkmate gives the output that FWR never enters to the "avoid" state, thereby confirming that the property passes. Table I shows the property validation results by Checkmate for the formal verification of FWR.

Table I: FWR PROPERTY VERIFICATION SUMMARY

| Functional Specification | Property Validation |
|---|---|
| P1 | Pass |
| P2 | Pass |

## V. CONCLUSION AND FUTURE WORK

In this paper formal verification of the FWR has been reported with Checkmate. To make it work we were needed some modifications in the Checkmate implementation. By making the above discussed changes in the Checkmate we are able to verify the two basic properties (safety property) of the FWR. In order to illustrate our work, only simple mixed-signal circuit is given in this paper. Nevertheless, the use of the modified Checkmate is applicable to complex mixed-signal circuits as well as to the non-linear analog circuits. A future research direction of this work will to be study its extension to mixed signal designs which come under the category of hybrid systems. More specifically, we intend to study this approach for mixed signal modules in a Sigma-Delta modulator block of an ADC system and the analog circuits having strong nonlinear behavior which are not so easy to verify formally due to their non-linear behavior.